\documentclass[prl,twocolumn,superscriptaddress]{revtex4-2}

\usepackage[utf8]{inputenc}
\usepackage{amsmath,amssymb}
\usepackage{braket}
\usepackage[colorlinks=true,linkcolor=blue,urlcolor=blue,citecolor=blue,anchorcolor=blue]{hyperref}
\usepackage[capitalise]{cleveref}
\Crefname{section}{Sec.}{Secs.}
\crefrangelabelformat{equation}{(#3#1#4--#5#2#6)}
\usepackage{tikz}
\usepackage{multirow}
\usepackage{amsfonts} 
\usepackage{diagbox}
\usepackage{subfigure}

\newcommand{\ie}{{\sl i.e.}}
\newcommand{\eg}{{\sl e.g.}}

\newcommand{\sect}[1]{{\it #1}.---}

\def\Xint#1{\mathchoice
   {\XXint\displaystyle\textstyle{#1}}%
   {\XXint\textstyle\scriptstyle{#1}}%
   {\XXint\scriptstyle\scriptscriptstyle{#1}}%
   {\XXint\scriptscriptstyle\scriptscriptstyle{#1}}%
   \!\int}
\def\XXint#1#2#3{{\setbox0=\hbox{$#1{#2#3}{\int}$}
     \vcenter{\hbox{$#2#3$}}\kern-.5\wd0}}
\def\pint{\Xint=}

\begin{document}
\title{Generalization of Weinberg's compositeness relations
}

 \newcommand{\UCAS}{School of Physical Sciences, University of Chinese Academy of Sciences (UCAS), Beijing 100049, China}
\newcommand{\USST}{College of Science, University of Shanghai for Science and Technology, Shanghai 200093, China}
\newcommand{\ITP}{CAS Key Laboratory of Theoretical Physics, Institute of Theoretical Physics,\\
Chinese Academy of Sciences, Beijing 100190, China }

 \author{Yan Li}\email{liyan175@mails.ucas.edu.cn }
 \affiliation{\UCAS}
 \author{Feng-Kun Guo}\email{fkguo@itp.ac.cn}
 \affiliation{\ITP}
 \affiliation{\UCAS}
 \author{Jin-Yi Pang} \email{jypang@usst.edu.cn, corresponding author}
 \affiliation{\USST}
 \author{Jia-Jun Wu} \email{wujiajun@ucas.ac.cn, corresponding author}
 \affiliation{\UCAS}

\begin{abstract}
    We generalize the time-honored Weinberg's compositeness relations by including the range corrections through considering a general form factor.
    In Weinberg's derivation, he considered the effective range expansion up to $\mathcal{O}(p^2)$ and made two additional approximations: neglecting the nonpole term in the Low equation and approximating the form factor by a constant.
    We lift the second approximation and work out an analytic expression for the form factor. 
    For a positive effective range, the form factor is of a single-pole form. 
    {An integral representation of the compositeness is obtained and is expected to have a smaller uncertainty than that derived from Weinberg's relations.}
    We also establish an exact relation between the wave function of a bound state and the phase of the scattering amplitude neglecting the nonpole term.
    The deuteron is analyzed as an example, and the formalism can be applied to other cases where range corrections are important.
\end{abstract}

\maketitle

\sect{Introduction}Deciding whether a particle is composite or elementary via low-energy scattering observables had been considered impossible until Weinberg proposed his relations~\cite{Weinberg:1965zz} that connect the scattering length $a$ and effective range $r$ with the compositeness of an $S$-wave shallow bound state.
For instance, the deuteron may be a superposition of a composite (molecular) state of two nucleons and a compact component (called an ``elementary-particle'' in the original publication~\cite{Weinberg:1965zz} to be distinct from the two-nucleon composite component).
The probability of finding the deuteron in the molecular state is called compositeness $X$.
The relations read~\cite{Weinberg:1965zz}
\begin{align}
    a &= -\frac{2X_W}{1+X_W}R + O(m_\pi^{-1}) \,,\label{eq:wrar1}\\
    r &= -\frac{1-X_W}{X_W}R + O(m_\pi^{-1}) \,,\label{eq:wrar2}
\end{align}
where $R={1}/{\sqrt{2\mu|E_B|}}$, $\mu$ is the reduced mass, $E_B$ is the binding energy, which is $E_B=-2.224575(9)\,$MeV~\cite{VanDerLeun:1982bhg}
for the deuteron case, and the $O(m_\pi^{-1})$ terms are due to neglecting the finite-range interactions.
We denote the compositeness from the above equations as $X_W$ in order to distinguish it from the one to be derived later.
Had the deuteron an appreciable 
compact
component,  Weinberg's relations would predict a large and negative $r$ and a small $a$, which clearly contradict the measured values~\cite{Klarsfeld:1984es} 
\begin{align}\label{eq:expar}
    a = -5.419(7)\,\text{fm} \,, \quad r = 1.766(8)\,\text{fm} \,.
\end{align}
So one can conclude that the deuteron is mostly a composite system of two nucleons.
The compositeness $X$, being a probability, should be in the range $[0,1]$. However, the $X_W$ value may go beyond that range. For instance, the above relations lead to a value of 1.68 for the deuteron, indicating a sizeable correction.
Besides the deuteron, Weinberg's relations and their extensions \cite{Baru:2003qq,Gamermann:2009uq,Baru:2010ww,Hanhart:2011jz,Hyodo:2011qc,Aceti:2012dd,Hyodo:2013iga,Hyodo:2013nka,Sekihara:2014kya,Hanhart:2014ssa,Guo:2015daa,Sekihara:2015gvw,Kamiya:2015aea,Xiao:2016dsx,Xiao:2016wbs,Kang:2016ezb,Sekihara:2016xnq,Kamiya:2016oao,Guo:2016wpy,Oller:2017alp,Kamiya:2017hni,Bruns:2019xgo,Matuschek:2020gqe} have been widely used to study many near-threshold hadrons (see Ref.~\cite{Guo:2017jvc} for a review).

On the other hand, the wave function of the deuteron has been studied by hundreds of works (see Ref.~\cite{Zhaba:2017syr} for a review).
Mostly constrained by the scattering phase shift, various models can predict wave functions with similar properties at long distances, suggesting a potential model-independent relationship between the two terms.
In fact, such a relation exists formally for purely local potentials in the context of the inverse scattering problem (see, \eg, Ref.~\cite{Chadan:1989Inverse}).
In the derivation of Weinberg's relations, the form factor of the deuteron, which encodes its coupling to the proton and neutron, is approximated by a constant (see below), which greatly limits the behavior of the wave function and, therefore, precludes a detailed study on it.

In this Letter, we will generalize the compositeness relations by lifting the constant-form-factor approximation completely. Consequently, {the uncertainty of compositeness is considerably reduced, and} a relation between the wave function and the scattering phase shift follows. The obtained compositeness for a bound state will always be in the range $[0,1]$. 
We will also derive an analytic expression for the form factor considering the effective range expansion (ERE).

\sect{Derivation}
{We restrict our discussion to the near-threshold region and thus consider only the $S$-wave interaction with a} Hamiltonian $\hat{H}=\hat{H}_0+\hat{V}$. Its half-shell $T$ matrix 
\begin{align}
    T_{p,k}:=\braket{p|\hat{T}(h_k+i\varepsilon)|k},
\end{align} 
with $p$ the momentum of one constituent particle in the center-of-mass frame of the two-body system, the normalization $\braket{p|k}={(2\pi)^3}\delta(p-k)/{p^2}$, and the kinetic energy $h_k={k^2}/{(2\mu)}$, is constrained by the Low equation~\cite{Weinberg:1965zz}
\begin{align}\label{eq:loweq}
    T_{p,k} = V_{p,k} + \frac{g(p)\, g^*(k)}{h_k-E_B } + \int_0^\infty \frac{q^2dq}{(2\pi)^3} \frac{T_{p,q}T_{k,q}^*}{h_k+i\varepsilon-h_q} \,,
\end{align}
where $V_{p,k}:=\braket{p|\hat{V}|k}$, $E_B<0$ is the binding energy of the bound state $\ket{B}$, and $g(p):=\braket{p|\hat{V}|B}$ is the form factor.

For a shallow bound state with $|E_B|\ll h_\Lambda := \Lambda^2/(2\mu)$, where $\Lambda$ is a hard momentum scale,
the second term on the right-hand side of \cref{eq:loweq} is enhanced by $(h_k-E_B)^{-1}$ in the low-momentum regime, where $k\ll \Lambda$ compared with the nonpole term $V_{p,k}$.
Then Weinberg made two approximations.
The first is to ignore the nonpole term $V_{p,k}$ in the Low equation.
The second is to replace the form factor $g(p)$ by a constant $g$, amounting to keeping only the leading-order term in a nonrelativistic expansion.
In the following, we adopt only the first one,
{i.e., neglecting the nonpole term, which should work well in the near-threshold region dominated by the pole. Note that a nonpole term of the form that can be generated by redefining $g(k)$  is already accounted for by the second term of Eq.~\eqref{eq:loweq}. }

{
With a separable ansatz, 
\begin{align}\label{eq:ansatz}
    T_{p,k} = t_k\, g(p)\, g^*(k) \,,
\end{align}
the Low equation can be solved (details can be found in the Supplemental Material):
\begin{align}\label{eq:ansatz_sol}
    T_{p,k} = \frac{1}{1-F(h_k)}\,\frac{g(p)\, g^*(k)}{h_k-E_B} \,,
\end{align}
with
\begin{align}\label{eq:FW}
    F(W) :=   \int_0^\infty \!\! \frac{q^2dq}{(2\pi)^3} \frac{(W-E_B)\,|g(q)|^2}{(h_q-E_B)^2(W-h_q)} \,.
\end{align}
The solution is obtained without the so-called Castillejo-Dalitz-Dyson  zeros~\cite{Castillejo:1955ed} as Weinberg did (for a discussion of the impact of such zeros on the compositeness, see Refs.~\cite{Baru:2010ww,Hanhart:2011jz,Kang:2016jxw}).
Interestingly, the compositeness $X$ shows up as
\begin{align}\label{eq:Finfty}
    F(\infty) &= \int_0^\infty \frac{q^2dq}{(2\pi)^3} \frac{|\braket{q|\hat{V}|B}|^2}{(h_q-E_B)^2} = \int_0^\infty \frac{q^2dq}{(2\pi)^3}\,|\braket{q|B}|^2 \nonumber\\
    &= X \,,
\end{align}
where the Schr\"odinger equation $\langle q | (h_q + \hat V) |B\rangle = E_B \langle q|B\rangle$ has been used.
The integral in \cref{eq:FW} can be solved in a closed form if $g(q)$ is approximated by a constant as done in Ref.~\cite{Weinberg:1965zz}.

One can define, with the convention $\delta_B(0)=0$,
\begin{align}\label{eq:argdelta}
    \delta_B(E=h_p):=\arg T_{p,p}=-\arg\left(1-F(E+i\varepsilon)\right)\,,
\end{align}
where $\delta_B$ means the phase of the on-shell $T$ matrix $T_{p,p}=t_p |g(p)|^2 $} 
and differs from the full phase shift $\delta$ by having neglected the nonpole term in Eq.~\eqref{eq:loweq} (for convenience, we take the convention $\delta(0)=0$, the same as that for $\delta_B$).
Nevertheless, we have $\delta_B\approx\delta$ in the low-momentum regime where the near-threshold pole dominates and the full on-shell $T$ matrix ($\propto e^{i\,\delta}\sin\delta$) is well approximated.
Furthermore, one can show 
\begin{align}\label{eq:resdelta}
    -\pi\leq\delta_B\leq 0
\end{align}
by noting 
$F(0)\leq 0$ and $\mathrm{Im}\,F(E+i\varepsilon)\leq 0$ for $E\geq 0$, and $\delta_B(\infty) = 0$ unless $X=1$.

{
One can work out a dispersive representation of $F(W)$ (see the Supplemental Material for the derivations),
}
\begin{align}\label{eq:FWdisp}
    F(W) = 1 - \exp\left(\frac{W-E_B}{\pi}\int_0^\infty dE \frac{-\delta_B(E)}{(E-W)(E-E_B)}\right) \,,
\end{align}
which, combined with \cref{eq:Finfty}, leads to {an integral representation of $X$ in terms of the low-energy observable $\delta_B$, which is the phase shift under the approximation neglecting $V_{p,k}$,}
\begin{align}\label{eq:Zdisp}
    X = 1 - \exp\left(\frac{1}{\pi}\int_0^\infty dE \frac{\delta_B(E)}{E-E_B}\right) \,.
\end{align}
The above formula presents a generalization of Weinberg's treatment by lifting the leading-order nonrelativistic approximation to the form factor (an expression of the compositeness in terms of $T$-matrix given by \cref{eq:loweq} can be found in Refs.~\cite{Hyodo:2011qc,Hyodo:2013nka}). It ensures that $X$ computed in this way cannot be larger than 1; furthermore, since $\delta_B \leq 0$, we also have $X\geq 0$. Thus, $X\in [0,1]$ is ensured, in contrast to that computed from Eqs.~\eqref{eq:wrar1} and \eqref{eq:wrar2}.

The form factor can also be constructed from $\delta_B$ by noticing
\begin{align}\label{eq:imF}
    \mathrm{Im}\,F(h_p+i\varepsilon)=-\frac{\pi\,p \mu}{(2\pi)^3}\frac{|g(p)|^2}{h_p-E_B} \,.
\end{align}
Working out the imaginary part of \cref{eq:FWdisp} and comparing it with \cref{eq:imF}, one finds
\begin{align}\label{eq:gp2}
    |g(p)|^2=&\, -\frac{(2\pi)^3}{\pi\,p \mu}(h_p-E_B)\,\sin\delta_B(E) \nonumber\\
    & \times\exp\left[\frac{h_p-E_B}{\pi}\,\pint_0^\infty dE \frac{-\delta_B(E)}{(E-h_p)(E-E_B)} \right] \,,
\end{align}
where $\pint$ denotes the principal value integral.

The form factor $g(p)$ is related to the radial wave function in the momentum space, $\tilde{u}(p)$, and that in the position space, $u(r)$, as follows:
\begin{align}
    \tilde{u}(p)&=\frac{p}{(2\pi)^{3/2}}\frac{g(p)}{h_p-E_B} \,,\label{eq:uu1}\\
    u(r)&=\frac{4\pi}{(2\pi)^{3/2}} \int_0^\infty dp\,\,\tilde{u}(p)\,\sin(p\,r)\,,\label{eq:uu2}
\end{align}
These wave functions are normalized as
\begin{align}
    \int_0^\infty dp\,|\tilde{u}(p)|^2 =\int_0^\infty dr\,|u(r)|^2=X \,.
\end{align}
We note that \cref{eq:gp2} does not determine the phase of $g(p)$.
However, if the system respects time reversal symmetry and the bound state is not degenerated, $u(r)$, $\tilde{u}(p)$ and $g(p)$ can all be made real.

Finally, with the convention in the current Letter, we can write the ERE,
\begin{align}\label{eq:p}
    p \cot \delta_B \approx -\frac{8\pi^2}{\mu}\text{Re}T^{-1}(h_p) 
    = \frac{1}{a} + \frac{r}{2}p^2 + \mathcal{O}(p^4),
\end{align}
where the $\approx$ is used to remind us that the nonpole term $V_{p,k}$ has been neglected from the Low equation.

\sect{Discussion}Equation~\eqref{eq:Zdisp} is a generalization of Weinberg's relations.
{It is expected to have a smaller uncertainty, of $\mathcal{O}(\Lambda^{-2})$, than that of Weinberg's relations, i.e. $\mathcal{O}(\Lambda^{-1})$ with $\Lambda\sim m_\pi$ for the deuteron case (for discussions of the uncertainty of Weinberg's relations, see, \eg, Refs.~\cite{Kamiya:2016oao,Kamiya:2017hni}). Here the small dimensionless quantity for estimating uncertainties needs to be understood as $1/(R\Lambda)$, and we neglect $R$ for simplicity.
Typically momenta appear in a square form in the potential. 
Therefore we expect the ignored $V_{p,k}$ only brings an uncertainty of $\mathcal{O}{(\Lambda^{-2})}$.
The reason that Weinberg's relations have an uncertainty of $\mathcal{O}{(\Lambda^{-1})}$ is that the constant-form-factor approximation, $g(p)=g_0$, has been applied to the estimation of $X$.
Although this approximation neglects the $\mathcal{O}{(\Lambda^{-2})}$ terms of $g^2(p)$, whose $\Lambda$-dependence reads $g^2(p)=g_0^2+\frac{p^2}{\Lambda^2}\tilde{g}^2(\frac{p^2}{\Lambda^2})$, the resulting uncertainty is
\begin{align}
    \Delta X  = \frac{1}{\Lambda^2} \int_0^\infty \frac{q^2dq}{(2\pi)^3} \frac{q^2\,\tilde{g}^2(\frac{q^2}{\Lambda^2})}{(h_q-E_B)^2} \sim \mathcal{O}(\Lambda^{-1}) \,,
    \label{eq:error}
\end{align}
where the integral is linearly divergent when $\Lambda\to\infty$ and needs to be cut at the hard scale $\Lambda$.
As the derivation of \cref{eq:Zdisp} does not rely on this approximation, we expect the uncertainty is of $\mathcal{O}{((R\Lambda)^{-2})}$.
}

In fact when we take the ERE up to $\mathcal{O}(p^2)$ for $\delta_B$ as given in \cref{eq:p}, an analytic model-independent, {based on the separable ansatz in Eq.~\eqref{eq:ansatz},} expression (the form factor was modeled in a Gaussian form in, e.g., Ref.~\cite{Faessler:2007gv} and with a hard cutoff in, e.g., Ref.~\cite{Gamermann:2009uq}) can be worked out for the form factor $g^2(p)$ from \cref{eq:gp2}, 
\begin{align}\label{eq:gp2_2}
    g^2(p)= \frac{8\pi^2}{\mu^2 R}\times\begin{cases}
    t + \mathcal{O}(p^4) & t\in[0,1] \\
    \frac{a^2}{R^2} \frac1{1 + (a+R)^2 p^2 }
    + \mathcal{O}(p^4)     & t>1 \\
    \end{cases},
\end{align}
where $t := 1/\sqrt{1+2r/a} = -a/(a+2R)$, which is just $X_W$ using Eqs.~\eqref{eq:wrar1} and \eqref{eq:wrar2}, and we have expressed $r$ in terms of $r = 2R(a+R)/a$ in the second line. 
The above expression is exact for ERE up to $\mathcal{O}(p^2)$, and  $\mathcal{O}(p^4)$ therein denotes that higher order terms have been neglected in the ERE. 

One sees that the form factor is a constant for $X_W\in[0,1]$, i.e.,
\begin{align}
  a\in [-R,0],\quad r\leq 0,
  \label{eq:arrange}
\end{align}
and the relation between $X_W$ and the coupling constant in Ref.~\cite{Weinberg:1965zz} is reproduced in this case.

However, when $X_W>1$, which corresponds to $a\in(-2R, -R)$ and $r\in (0,R)$,
$g^2(p)$ contains $\mathcal{O}(p^2)$ terms and is of a single-pole form.
Since the form factor enters the $T$-matrix through the Low equation~\eqref{eq:loweq}, such $\mathcal{O}(p^2)$ terms would contribute to the effective range, and need to be taken into account consistently up to $\mathcal{O}(p^2)$.
However, it was neglected in the original treatment of Ref.~\cite{Weinberg:1965zz}; then for $r>0$ the value of $X_W$ solved using a constant form factor (or coupling) is larger than 1, as is the case for the deuteron ($X_W=1.68$), and loses its direct interpretation as a probability.
On the contrary, the compositeness given in \cref{eq:Zdisp}, which is the exact solution of the Low equation with the nonpole term $V_{p,k}$ neglected, will be exactly $X = 1$
because $\delta_B(\infty)=-\pi$ as can be seen from Eq.~\eqref{eq:p}.
Thus, we conclude that the compositeness of the deuteron is 1. The same also applies to the $D_{s0}^*(2317)$ and $D_{s1}(2460)$ as computed by the RQCD Collaboration using lattice quantum chromodynamics in Ref.~\cite{Bali:2017pdv}, where the isoscalar $DK$ and $D^*K$ effective ranges are positive and the authors obtained larger-than-1 values for the compositeness using Eqs.~\eqref{eq:wrar1} and \eqref{eq:wrar2}. 
For comparison, the $DK$ effective range obtained by the Hadron Spectrum Collaboration is negative, and the extracted compositeness close to 1 using the same relations is valid~\cite{Cheung:2020mql}.
From the above discussion, considering the binding energy ranging from 22~MeV to 60~MeV at the two pion masses therein~\cite{Cheung:2020mql}, the systematic uncertainty of $X$ from using Weinberg's relations should be about $(R\Lambda)^{-2} \in [0.03, 0.07]$, which is much less than $(R\Lambda)^{-1}\in [0.16, 0.27]$ that one would naively assume. Here $\Lambda\sim m_\rho$ is estimated from the lightest meson exchanged between $D^{(*)}K$.

Since Wigner's causality inequality constrains the effective range $r$ to be negative semi-definite for a zero-range interaction (see, e.g., Refs.~\cite{Bohm:book,Matuschek:2020gqe}), a positive $r$ implies a sizable range correction. 
This is reflected by the $p^2$ dependence in \cref{eq:gp2_2}. 
For the deuteron case, one has $|a+R| \approx 1.1$~fm, at the order of the inverse of the pion mass.

There is a crucial difference between a negative and a positive $r$ (see also Refs.~\cite{Hyodo:2013iga,Hanhart:2014ssa,Matuschek:2020gqe} for related discussions). 
The $T$-matrix has two poles in the complex momentum plane using ERE up to a nonvanishing effective range term, which are located at $p_-=i/R$ and $p_+ = -i/(R+a)$. Here we have chosen to express the quantities in terms of $a$ and $R$, and the effective range is $r=2R(R+a)/a$. $a$ must be negative in order to have a bound state pole.
For $a\in(-R,0)$, $p_-$ is a bound state pole, and $p_+$ is a remote virtual state pole; in this case, $r<0$, and, correspondingly, the form factor is just a constant up to $\mathcal{O}(p^2)$.
However, for $a\in(-2R,-R)$, $p_+$ becomes a bound state pole as well; in this case, $r\in(0,R)$. 
Although this pole is spurious, it effectively resums range corrections (see Ref.~\cite{Hanhart:2007wa} for discussions of range corrections in two-photon decays of hadronic molecules), and the form factor in Eq.~\eqref{eq:gp2_2} is indeed proportional to $1/(p^2 - p_+^2)$.
For $a<-2R$, the roles of $p_+$ and $p_-$ are interchanged, and the above discussions still apply. For the fine tuning case $a\approx -2R$, the poles are close to each other and the Low equation needs to be modified.

The result of \cref{eq:Zdisp} for $X$ 
has an uncertainty because $\delta_B$ is obtained by neglecting the nonpole term $V_{p,k}$, which could have a sizeable contribution in the high-momentum range.
Thus, we introduce
\begin{align}
    &\delta_B(E,\underline\Lambda)=\begin{cases}
    \delta_B(E) & E\leq h_\Lambda \\
    0 & E>h_\Lambda \\
    \end{cases}\,, \nonumber\\
    &\delta_B(E,\overline\Lambda)=\begin{cases}
        \delta_B(E) & E\leq h_\Lambda \\
        -\pi & E>h_\Lambda \\
    \end{cases} \,,
\end{align}
with $h_\Lambda = \Lambda^2/(2\mu)$
so that
\begin{align}
    -\pi\leq\delta_B(E,\overline\Lambda)\leq\delta_B(E)\leq\delta_B(E,\underline\Lambda)\leq 0 \,.
\end{align}
The corresponding compositenesses, form factors and wave functions will be denoted in a similar way.
For the compositeness, one has 
\begin{align}\label{eq:ncp}
    X(\underline\Lambda) &= 1 - \exp\left(\frac{1}{\pi}\int_0^{h_\Lambda} dE \frac{\delta_B(E)}{E-E_B}\right)  \nonumber\\
    &\leq X \leq 1 = X(\overline\Lambda) \,.
\end{align}
For the form factor, one can show
\begin{align}
    &g^2(p,\underline\Lambda)\leq g^2(p)\leq g^2(p,\overline\Lambda) \,,\text{ when } p\leq \Lambda\label{eq:gg1}\,,\\
    &g^2(p,\underline\Lambda)=g^2(p,\overline\Lambda)=0 \,,\quad\;\,\text{ when } p>\Lambda \,.\label{eq:gg2}
\end{align}
Thus, with a larger $\Lambda$, one has tighter bounds on $X$ which, on the other hand, bear a larger uncertainty.
For the wave function in the momentum space, there are similar relations. For the position space, however, there are not.

For a shallow bound state, the ${p}/{(h_p-E_B)}$ factor in \cref{eq:uu1} peaks sharply around $p=\sqrt{2\mu|E_B|}$.
Once $\Lambda$ is chosen beyond the peaking range, $\tilde{u}(p)$ will be largely determined by the long-distance physics.
When we go to the position space, the short-distance (small-$r$) part of $u(r)$ receives little contribution from the small-$p$ part of $\tilde{u}(p)$ as a consequence of the uncertainty principle.
To be more concrete, because of the $\sin(p\,r)$ term in \cref{eq:uu2}, $\tilde{u}(p)$ contributes to $u(r)$ only when $p\gtrsim {\pi}/{(2r)}$.
This reflects the inability to probe the short-distance structure of a shallow bound state using the information of low-energy scattering.
For the deuteron, for instance, it is impossible to distinguish the compact nucleon-nucleon component, \ie, the component of the small-$r$ part of $u(r)$, from a possible ``elementary-particle" core using only the low-energy nucleon-nucleon scattering.

\sect{Analysis of deuteron}As discussed before, our prediction for the compositeness of the deuteron is simply $100\%$.
The uncertainty of this value is determined by  how well $\delta_B$ approximates the genuine phase shift.
{
As argued above around \cref{eq:error}, 
for the deuteron,
while Weinberg's relations would predict $X=1.68+\mathcal{O}((Rm_\pi)^{-1}\simeq 0.3)$, we have $X=1+\mathcal{O}(0.3^2=0.09)$.
}

\begin{figure*}[tbp]
    \centering
        \includegraphics[width=0.32\textwidth]{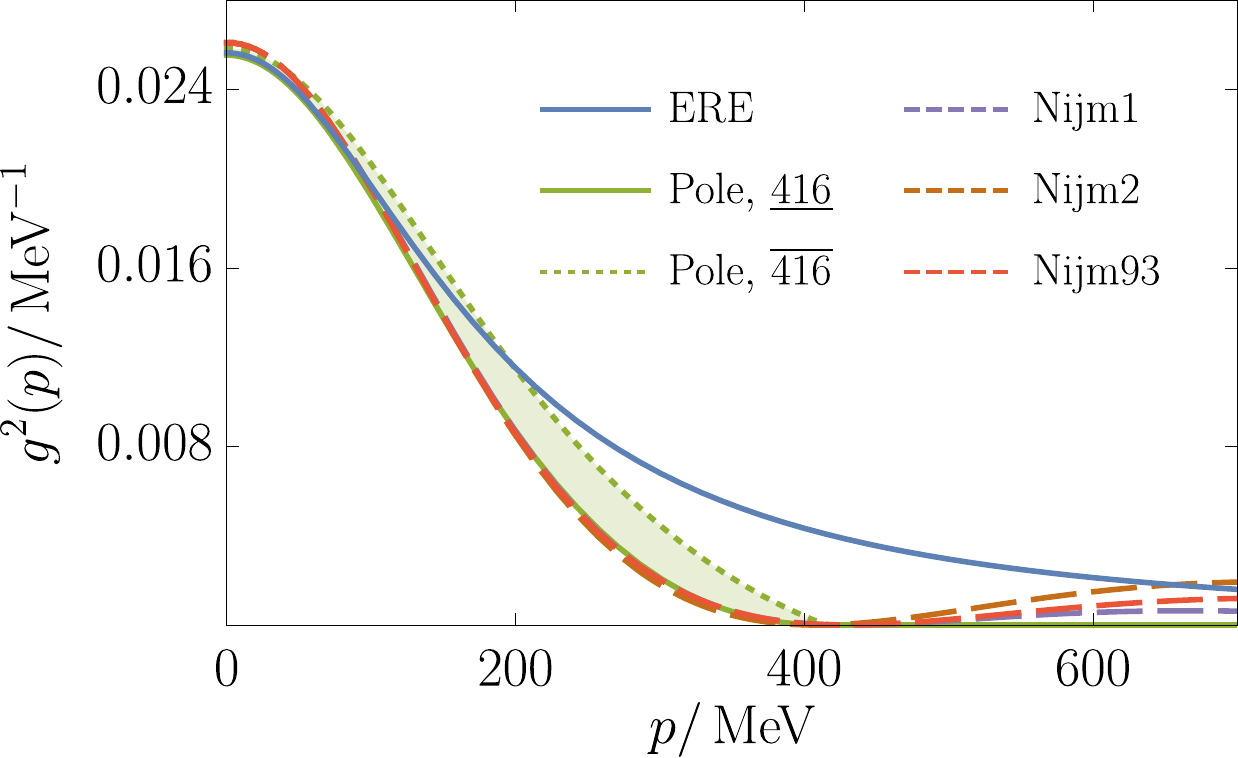} \hfill
        \includegraphics[width=0.32\textwidth]{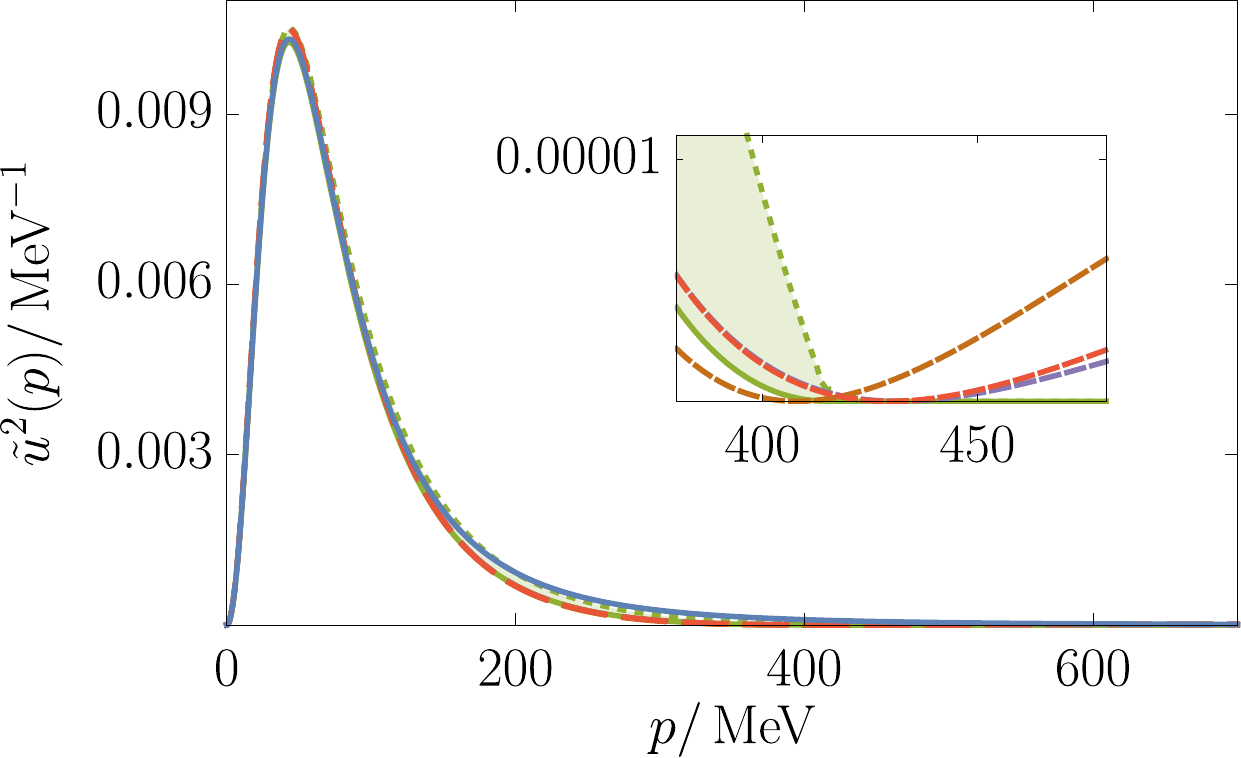} \hfill
        \includegraphics[width=0.32\textwidth]{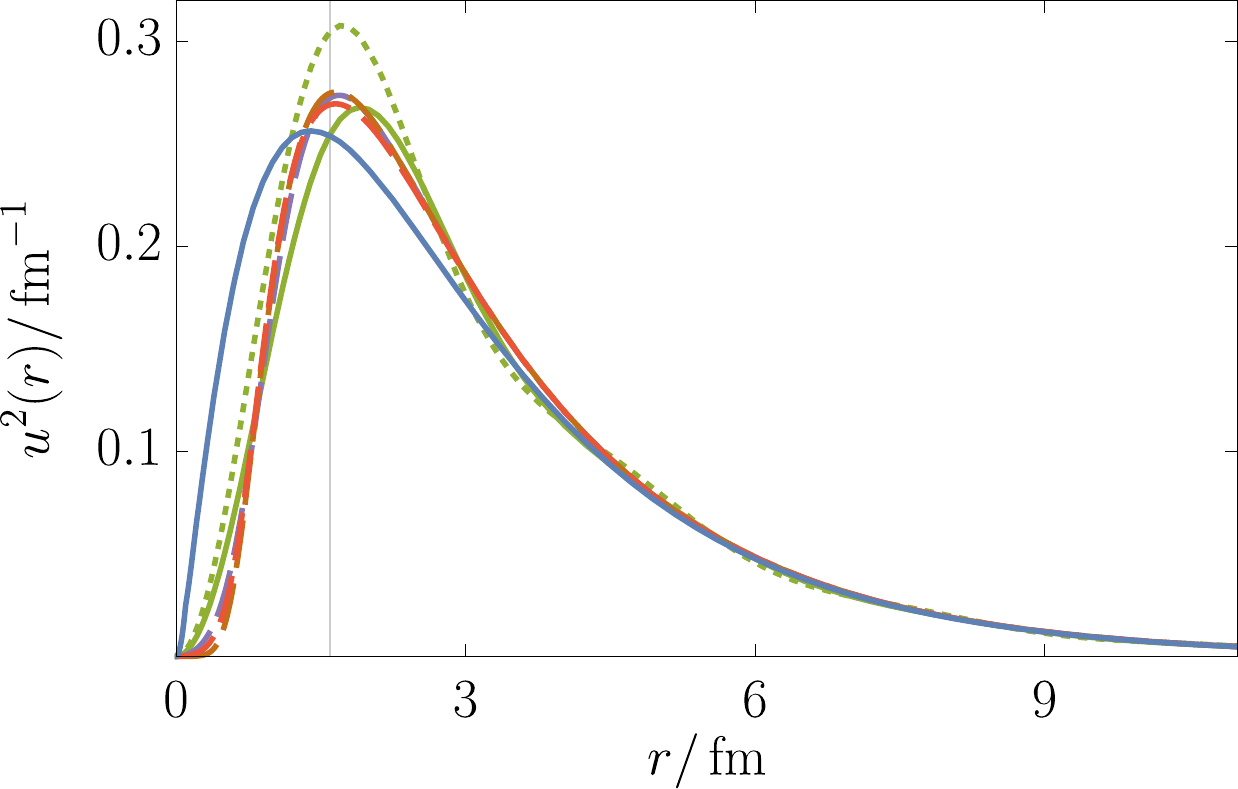}        
    \caption{
        The form factors (left) and wave functions in the momentum space (middle) and position space (right) for various cases: obtained from $\delta_B=\delta_{\text{ERE}}$ (blue), obtained from $\delta_B=\delta_{\text{Pole}}$ from \cref{eq:real} with a cutoff $416\,$MeV (green), and using three Nijmegen models (dashed).
        The shaded area indicates that they are the bounds for $g^2(p)$ and $\tilde{u}^2(p)$.
        The vertical gray line in the right plot denotes $1.6\,$fm.
    }\label{fig:guu}
\end{figure*}

We plot the form factor and wave functions of the deuteron in \cref{fig:guu}, where we also include the results from three famous Nijmegen models~\cite{Stoks:1994wp,nnonline} (the references provide wave functions both in position and momentum spaces directly, and the form factors are obtained from \cref{eq:uu1}).
The three Nijmegen models give $X=(94.246$--$94.365)\%$ (the remaining $1-X$ part comes from the $D$-wave component).
It is interesting to find that the ERE phase shift produces a form factor that drops quickly, as given in the second line of \cref{eq:gp2_2}, showing the importance of the range corrections neglected in Weinberg's relations.
Compared to Nijmegen models, the discrepancy of the form factor from ERE becomes larger at higher momenta unsurprisingly.
However, thanks to the ${p}/{(h_p-E_B)}$ factor in \cref{eq:uu1}, the wave function $\tilde{u}^2(p)$ drops faster than $g^2(p)$, so the discrepancy becomes hardly visible for $\tilde{u}^2(p)$ even in the high momentum range.
For the wave function in position space, the discrepancy is mainly in the small-$r$ range as expected.
The deuteron wave function has been computed using many other methods; \eg, Ref.~\cite{Zhaba:2017syr} compared wave functions from the Nijmegen group potentials \cite{Stoks:1994wp} and Argonne v18 potential \cite{Wiringa:1994wb}, Ref.~\cite{Epelbaum:2014efa}  compared their results with the results from the Idaho (500) N$^3$LO potential of Ref.~\cite{Entem:2003ft}, the N$^3$LO (550/600) potential of Ref.~\cite{Epelbaum:2004fk} and the CD-Bonn potential \cite{Machleidt:2000ge}, and  Ref.~\cite{Nogga:2005fv} discussed wave functions in an EFT context.
Most of these wave functions share a quite similar large-$r$ behavior, and differ mainly in the $r\lesssim2\,$fm range. 

We can even go beyond the ERE by using a parametrization fitted to the experimental phase shifts up to a higher momentum.
We adopt the following parametrization used in Ref.~\cite{Babenko:2005qp}, where the authors refer to it as the pole approximation:
\begin{align}\label{eq:real}
    p \cot \delta_{\text{Pole}} = \frac{1}{a} + \frac{r}{2}p^2 + \frac{v_2\, p^4}{1-D\,p^2} ,
\end{align}
with $a=-5.4030\,$fm, $r=1.7494\,$fm, $v_2 = 0.163\,\text{fm}^3$, and $D = 0.225526\,\text{fm}^2$.
When $p>D^{-1/2}=416\,$MeV, $\delta_{\text{Pole}}<-\pi$, violating the restriction \cref{eq:resdelta}.
So we have to introduce a cutoff $\Lambda=416\,$MeV, with which $X(\underline\Lambda)=92.8\%$,
{
consistent with $X=1+\mathcal{O}(0.09)$ given above.
We also note that, even with a much smaller $\Lambda$, \eg, one would have $X \geq X(\underline{m_\pi})=62\%$ is already larger than $50\%$, which reliably concludes the deuteron to be mostly composite.
}
The corresponding form factors and wave functions are included in \cref{fig:guu}.
It is clear from the $g^2(p)$ plots that $\delta_{\text{Pole}}$ is better than $\delta_{\text{ERE}}$ at approximating Nijmegen models.

\begin{figure}[tbp]
    \centering
    \includegraphics[width=\columnwidth]{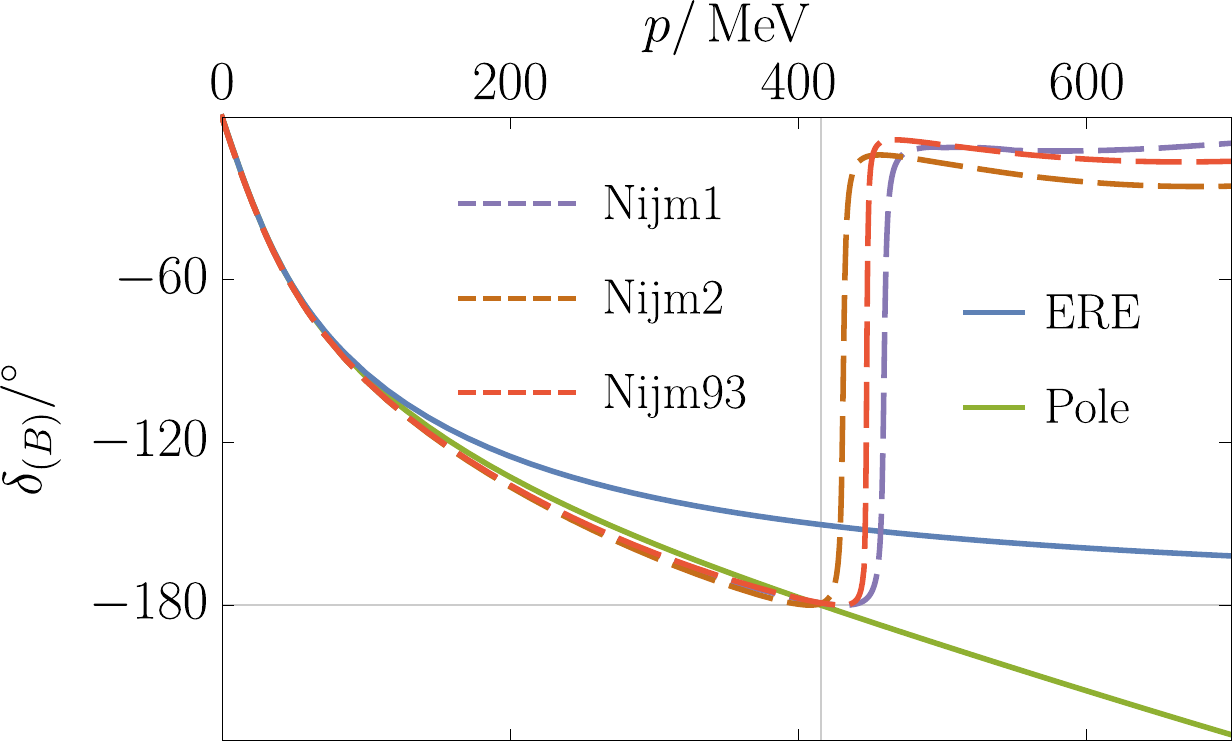}
    \caption{$\delta_B$ from ERE and the Nijmegen models and the phase shift from an empirical parameterization used in Ref.~\cite{Babenko:2005qp} (denoted by ``Pole'').
    The vertical gray line denotes $D^{-1/2}=416\,$MeV.
    }\label{fig:delta}
\end{figure}

Although it is hard to quantify the uncertainty of neglecting the nonpole term $V_{p,k}$ from the Low equation, such an approximation in fact works rather well up to a relatively high momentum for the deuteron case as can be seen from \cref{fig:delta}.
Here, the $\delta_B$ curves for the Nijmegen models and for the ERE are shown together with the empirical phase shift $\delta_{\text{Pole}}$.
It is clear that $\delta_B$ from the Nijmegen models approximate $\delta_{\text{Pole}}$ very well even up to the cutoff $416\,$MeV.
Note that although the nonpole term is neglected from the Low equation, the resulting $T$-matrix is more than simply a pole term.

\sect{Summary}In this Letter, we have generalized Weinberg's compositeness relations with a general form factor (and thus the range corrections are included),
which also builds an exact relation between the phase shift $\delta_B$, with the nonpole term neglected from the Low equation, and the bound state wave function.
For a shallow bound state, $\delta_B$ can approximate well the full phase shift in the low-momentum range, which goes up to more than 400~MeV for the deuteron case.
The compositeness derived from \cref{eq:Zdisp} is strictly within $[0,1]$ for a bound state even when the range corrections are important.
{It is expected to have a significantly smaller uncertainty than that from Weinberg's relations.}
An analytic expression for the form factor is obtained considering the ERE up to $\mathcal{O}(p^2)$. It is a constant for $a\in [-R,0]$, and $r\leq 0$, verifying Weinberg's approximation. 
However, if the effective range is positive, the form factor contains $\mathcal{O}(p^2)$ contributions in a single-pole form, and thus Weinberg's treatment is not self-consistent as it keeps only part of the $\mathcal{O}(p^2)$ contributions.
We then analyzed the deuteron as an example.
The range corrections are important in this famous case, as reflected in the strongly momentum-dependent form factor derived here.

{Given that many near-threshold states were observed in particular in the past two decades, the formalism will find its applications in such systems. Of particular importance is that the uncertainty of the extracted compositeness will be substantially reduced compared to that from Weinberg's relations.
The formalism can also be applied to other composite systems with short-range interactions beyond hadron physics.
}

\medskip
\begin{acknowledgments}
We would like to thank Christoph Hanhart, Tsung-Shung Harry Lee, De-Shan Yang, Ross D. Young, James M. Zanotti, and Bing-Song Zou for helpful comments and discussions.
This work is supported in part by the Chinese Academy of Sciences (CAS) under Grants No.~XDPB15, No.~XDB34030000 and No.~QYZDB-SSW-SYS013, by the National Natural Science Foundation of China (NSFC) under Grants No.~12125507, No.~11835015, No.~12047503 and No.~11961141012, and by the NSFC and the Deutsche Forschungsgemeinschaft (DFG, German Research Foundation) through funds provided to the Sino-German Collaborative Research Center ``Symmetries and the Emergence of Structure in QCD'' (NSFC Grant No. 12070131001, DFG Project-ID 196253076 -- TRR110),
by the Fundamental Research Funds for the Central Universities, and by the National
Key R$\&$D Program of China under Contract No. 2020YFA0406400.

\end{acknowledgments}

\bibliography{refsDeltaWave.bib}

\begin{thebibliography}{45}%
\makeatletter
\providecommand \@ifxundefined [1]{%
 \@ifx{#1\undefined}
}%
\providecommand \@ifnum [1]{%
 \ifnum #1\expandafter \@firstoftwo
 \else \expandafter \@secondoftwo
 \fi
}%
\providecommand \@ifx [1]{%
 \ifx #1\expandafter \@firstoftwo
 \else \expandafter \@secondoftwo
 \fi
}%
\providecommand \natexlab [1]{#1}%
\providecommand \enquote  [1]{``#1''}%
\providecommand \bibnamefont  [1]{#1}%
\providecommand \bibfnamefont [1]{#1}%
\providecommand \citenamefont [1]{#1}%
\providecommand \href@noop [0]{\@secondoftwo}%
\providecommand \href [0]{\begingroup \@sanitize@url \@href}%
\providecommand \@href[1]{\@@startlink{#1}\@@href}%
\providecommand \@@href[1]{\endgroup#1\@@endlink}%
\providecommand \@sanitize@url [0]{\catcode `\\12\catcode `\$12\catcode
  `\&12\catcode `\#12\catcode `\^12\catcode `\_12\catcode `\%12\relax}%
\providecommand \@@startlink[1]{}%
\providecommand \@@endlink[0]{}%
\providecommand \url  [0]{\begingroup\@sanitize@url \@url }%
\providecommand \@url [1]{\endgroup\@href {#1}{\urlprefix }}%
\providecommand \urlprefix  [0]{URL }%
\providecommand \Eprint [0]{\href }%
\providecommand \doibase [0]{https://doi.org/}%
\providecommand \selectlanguage [0]{\@gobble}%
\providecommand \bibinfo  [0]{\@secondoftwo}%
\providecommand \bibfield  [0]{\@secondoftwo}%
\providecommand \translation [1]{[#1]}%
\providecommand \BibitemOpen [0]{}%
\providecommand \bibitemStop [0]{}%
\providecommand \bibitemNoStop [0]{.\EOS\space}%
\providecommand \EOS [0]{\spacefactor3000\relax}%
\providecommand \BibitemShut  [1]{\csname bibitem#1\endcsname}%
\let\auto@bib@innerbib\@empty
\bibitem [{\citenamefont {Weinberg}(1965)}]{Weinberg:1965zz}%
  \BibitemOpen
  \bibfield  {author} {\bibinfo {author} {\bibfnamefont {S.}~\bibnamefont
  {Weinberg}},\ }\bibfield  {title} {\bibinfo {title} {{Evidence that the
  deuteron is not an elementary particle}},\ }\href
  {https://doi.org/10.1103/PhysRev.137.B672} {\bibfield  {journal} {\bibinfo
  {journal} {Phys. Rev.}\ }\textbf {\bibinfo {volume} {137}},\ \bibinfo {pages}
  {B672} (\bibinfo {year} {1965})}\BibitemShut {NoStop}%
\bibitem [{\citenamefont {Van Der~Leun}\ and\ \citenamefont
  {Alderliesten}(1982)}]{VanDerLeun:1982bhg}%
  \BibitemOpen
  \bibfield  {author} {\bibinfo {author} {\bibfnamefont {C.}~\bibnamefont {Van
  Der~Leun}}\ and\ \bibinfo {author} {\bibfnamefont {C.}~\bibnamefont
  {Alderliesten}},\ }\bibfield  {title} {\bibinfo {title} {{The deuteron
  binding energy}},\ }\href {https://doi.org/10.1016/0375-9474(82)90105-1}
  {\bibfield  {journal} {\bibinfo  {journal} {Nucl. Phys. A}\ }\textbf
  {\bibinfo {volume} {380}},\ \bibinfo {pages} {261} (\bibinfo {year}
  {1982})}\BibitemShut {NoStop}%
\bibitem [{\citenamefont {Klarsfeld}\ \emph {et~al.}(1984)\citenamefont
  {Klarsfeld}, \citenamefont {Martorell},\ and\ \citenamefont
  {Sprung}}]{Klarsfeld:1984es}%
  \BibitemOpen
  \bibfield  {author} {\bibinfo {author} {\bibfnamefont {S.}~\bibnamefont
  {Klarsfeld}}, \bibinfo {author} {\bibfnamefont {J.}~\bibnamefont
  {Martorell}},\ and\ \bibinfo {author} {\bibfnamefont {D.~W.~L.}\ \bibnamefont
  {Sprung}},\ }\bibfield  {title} {\bibinfo {title} {{Deuteron properties and
  the nucleon nucleon interaction}},\ }\href
  {https://doi.org/10.1088/0305-4616/10/2/008} {\bibfield  {journal} {\bibinfo
  {journal} {J. Phys. G}\ }\textbf {\bibinfo {volume} {10}},\ \bibinfo {pages}
  {165} (\bibinfo {year} {1984})}\BibitemShut {NoStop}%
\bibitem [{\citenamefont {Baru}\ \emph {et~al.}(2004)\citenamefont {Baru},
  \citenamefont {Haidenbauer}, \citenamefont {Hanhart}, \citenamefont
  {Kalashnikova},\ and\ \citenamefont {Kudryavtsev}}]{Baru:2003qq}%
  \BibitemOpen
  \bibfield  {author} {\bibinfo {author} {\bibfnamefont {V.}~\bibnamefont
  {Baru}}, \bibinfo {author} {\bibfnamefont {J.}~\bibnamefont {Haidenbauer}},
  \bibinfo {author} {\bibfnamefont {C.}~\bibnamefont {Hanhart}}, \bibinfo
  {author} {\bibfnamefont {Y.}~\bibnamefont {Kalashnikova}},\ and\ \bibinfo
  {author} {\bibfnamefont {A.~E.}\ \bibnamefont {Kudryavtsev}},\ }\bibfield
  {title} {\bibinfo {title} {{Evidence that the $a_0(980)$ and $f_0(980)$ are
  not elementary particles}},\ }\href
  {https://doi.org/10.1016/j.physletb.2004.01.088} {\bibfield  {journal}
  {\bibinfo  {journal} {Phys. Lett. B}\ }\textbf {\bibinfo {volume} {586}},\
  \bibinfo {pages} {53} (\bibinfo {year} {2004})},\ \Eprint
  {https://arxiv.org/abs/hep-ph/0308129} {arXiv:hep-ph/0308129} \BibitemShut
  {NoStop}%
\bibitem [{\citenamefont {Gamermann}\ \emph {et~al.}(2010)\citenamefont
  {Gamermann}, \citenamefont {Nieves}, \citenamefont {Oset},\ and\
  \citenamefont {Ruiz~Arriola}}]{Gamermann:2009uq}%
  \BibitemOpen
  \bibfield  {author} {\bibinfo {author} {\bibfnamefont {D.}~\bibnamefont
  {Gamermann}}, \bibinfo {author} {\bibfnamefont {J.}~\bibnamefont {Nieves}},
  \bibinfo {author} {\bibfnamefont {E.}~\bibnamefont {Oset}},\ and\ \bibinfo
  {author} {\bibfnamefont {E.}~\bibnamefont {Ruiz~Arriola}},\ }\bibfield
  {title} {\bibinfo {title} {{Couplings in coupled channels versus wave
  functions: application to the $X(3872)$ resonance}},\ }\href
  {https://doi.org/10.1103/PhysRevD.81.014029} {\bibfield  {journal} {\bibinfo
  {journal} {Phys. Rev. D}\ }\textbf {\bibinfo {volume} {81}},\ \bibinfo
  {pages} {014029} (\bibinfo {year} {2010})},\ \Eprint
  {https://arxiv.org/abs/0911.4407} {arXiv:0911.4407 [hep-ph]} \BibitemShut
  {NoStop}%
\bibitem [{\citenamefont {Baru}\ \emph {et~al.}(2010)\citenamefont {Baru},
  \citenamefont {Hanhart}, \citenamefont {Kalashnikova}, \citenamefont
  {Kudryavtsev},\ and\ \citenamefont {Nefediev}}]{Baru:2010ww}%
  \BibitemOpen
  \bibfield  {author} {\bibinfo {author} {\bibfnamefont {V.}~\bibnamefont
  {Baru}}, \bibinfo {author} {\bibfnamefont {C.}~\bibnamefont {Hanhart}},
  \bibinfo {author} {\bibfnamefont {Y.~S.}\ \bibnamefont {Kalashnikova}},
  \bibinfo {author} {\bibfnamefont {A.~E.}\ \bibnamefont {Kudryavtsev}},\ and\
  \bibinfo {author} {\bibfnamefont {A.~V.}\ \bibnamefont {Nefediev}},\
  }\bibfield  {title} {\bibinfo {title} {{Interplay of quark and meson degrees
  of freedom in a near-threshold resonance}},\ }\href
  {https://doi.org/10.1140/epja/i2010-10929-7} {\bibfield  {journal} {\bibinfo
  {journal} {Eur. Phys. J. A}\ }\textbf {\bibinfo {volume} {44}},\ \bibinfo
  {pages} {93} (\bibinfo {year} {2010})},\ \Eprint
  {https://arxiv.org/abs/1001.0369} {arXiv:1001.0369 [hep-ph]} \BibitemShut
  {NoStop}%
\bibitem [{\citenamefont {Hanhart}\ \emph {et~al.}(2011)\citenamefont
  {Hanhart}, \citenamefont {Kalashnikova},\ and\ \citenamefont
  {Nefediev}}]{Hanhart:2011jz}%
  \BibitemOpen
  \bibfield  {author} {\bibinfo {author} {\bibfnamefont {C.}~\bibnamefont
  {Hanhart}}, \bibinfo {author} {\bibfnamefont {Y.~S.}\ \bibnamefont
  {Kalashnikova}},\ and\ \bibinfo {author} {\bibfnamefont {A.~V.}\ \bibnamefont
  {Nefediev}},\ }\bibfield  {title} {\bibinfo {title} {{Interplay of quark and
  meson degrees of freedom in a near-threshold resonance: multi-channel
  case}},\ }\href {https://doi.org/10.1140/epja/i2011-11101-9} {\bibfield
  {journal} {\bibinfo  {journal} {Eur. Phys. J. A}\ }\textbf {\bibinfo {volume}
  {47}},\ \bibinfo {pages} {101} (\bibinfo {year} {2011})},\ \Eprint
  {https://arxiv.org/abs/1106.1185} {arXiv:1106.1185 [hep-ph]} \BibitemShut
  {NoStop}%
\bibitem [{\citenamefont {Hyodo}\ \emph {et~al.}(2012)\citenamefont {Hyodo},
  \citenamefont {Jido},\ and\ \citenamefont {Hosaka}}]{Hyodo:2011qc}%
  \BibitemOpen
  \bibfield  {author} {\bibinfo {author} {\bibfnamefont {T.}~\bibnamefont
  {Hyodo}}, \bibinfo {author} {\bibfnamefont {D.}~\bibnamefont {Jido}},\ and\
  \bibinfo {author} {\bibfnamefont {A.}~\bibnamefont {Hosaka}},\ }\bibfield
  {title} {\bibinfo {title} {Compositeness of dynamically generated states in a
  chiral unitary approach},\ }\href
  {https://doi.org/10.1103/PhysRevC.85.015201} {\bibfield  {journal} {\bibinfo
  {journal} {Phys. Rev. C}\ }\textbf {\bibinfo {volume} {85}},\ \bibinfo
  {pages} {015201} (\bibinfo {year} {2012})},\ \Eprint
  {https://arxiv.org/abs/1108.5524} {arXiv:1108.5524} \BibitemShut {NoStop}%
\bibitem [{\citenamefont {Aceti}\ and\ \citenamefont
  {Oset}(2012)}]{Aceti:2012dd}%
  \BibitemOpen
  \bibfield  {author} {\bibinfo {author} {\bibfnamefont {F.}~\bibnamefont
  {Aceti}}\ and\ \bibinfo {author} {\bibfnamefont {E.}~\bibnamefont {Oset}},\
  }\bibfield  {title} {\bibinfo {title} {{Wave functions of composite hadron
  states and relationship to couplings of scattering amplitudes for general
  partial waves}},\ }\href {https://doi.org/10.1103/PhysRevD.86.014012}
  {\bibfield  {journal} {\bibinfo  {journal} {Phys. Rev. D}\ }\textbf {\bibinfo
  {volume} {86}},\ \bibinfo {pages} {014012} (\bibinfo {year} {2012})},\
  \Eprint {https://arxiv.org/abs/1202.4607} {arXiv:1202.4607 [hep-ph]}
  \BibitemShut {NoStop}%
\bibitem [{\citenamefont {Hyodo}(2013{\natexlab{a}})}]{Hyodo:2013iga}%
  \BibitemOpen
  \bibfield  {author} {\bibinfo {author} {\bibfnamefont {T.}~\bibnamefont
  {Hyodo}},\ }\bibfield  {title} {\bibinfo {title} {{Structure of
  near-threshold $s$-wave resonances}},\ }\href
  {https://doi.org/10.1103/PhysRevLett.111.132002} {\bibfield  {journal}
  {\bibinfo  {journal} {Phys. Rev. Lett.}\ }\textbf {\bibinfo {volume} {111}},\
  \bibinfo {pages} {132002} (\bibinfo {year} {2013}{\natexlab{a}})},\ \Eprint
  {https://arxiv.org/abs/1305.1999} {arXiv:1305.1999 [hep-ph]} \BibitemShut
  {NoStop}%
\bibitem [{\citenamefont {Hyodo}(2013{\natexlab{b}})}]{Hyodo:2013nka}%
  \BibitemOpen
  \bibfield  {author} {\bibinfo {author} {\bibfnamefont {T.}~\bibnamefont
  {Hyodo}},\ }\bibfield  {title} {\bibinfo {title} {Structure and compositeness
  of hadron resonances},\ }\href {https://doi.org/10.1142/S0217751X13300457}
  {\bibfield  {journal} {\bibinfo  {journal} {Int. J. Mod. Phys. A}\ }\textbf
  {\bibinfo {volume} {28}},\ \bibinfo {pages} {1330045} (\bibinfo {year}
  {2013}{\natexlab{b}})},\ \Eprint {https://arxiv.org/abs/1310.1176}
  {arXiv:1310.1176} \BibitemShut {NoStop}%
\bibitem [{\citenamefont {Sekihara}\ \emph {et~al.}(2015)\citenamefont
  {Sekihara}, \citenamefont {Hyodo},\ and\ \citenamefont
  {Jido}}]{Sekihara:2014kya}%
  \BibitemOpen
  \bibfield  {author} {\bibinfo {author} {\bibfnamefont {T.}~\bibnamefont
  {Sekihara}}, \bibinfo {author} {\bibfnamefont {T.}~\bibnamefont {Hyodo}},\
  and\ \bibinfo {author} {\bibfnamefont {D.}~\bibnamefont {Jido}},\ }\bibfield
  {title} {\bibinfo {title} {Comprehensive analysis of the wave function of a
  hadronic resonance and its compositeness},\ }\href
  {https://doi.org/10.1093/ptep/ptv081} {\bibfield  {journal} {\bibinfo
  {journal} {Progress of Theoretical and Experimental Physics}\ }\textbf
  {\bibinfo {volume} {2015}},\ \bibinfo {pages} {63D04} (\bibinfo {year}
  {2015})},\ \Eprint {https://arxiv.org/abs/1411.2308} {arXiv:1411.2308}
  \BibitemShut {NoStop}%
\bibitem [{\citenamefont {Hanhart}\ \emph {et~al.}(2014)\citenamefont
  {Hanhart}, \citenamefont {Pel{\'a}ez},\ and\ \citenamefont
  {R{\'i}os}}]{Hanhart:2014ssa}%
  \BibitemOpen
  \bibfield  {author} {\bibinfo {author} {\bibfnamefont {C.}~\bibnamefont
  {Hanhart}}, \bibinfo {author} {\bibfnamefont {J.~R.}\ \bibnamefont
  {Pel{\'a}ez}},\ and\ \bibinfo {author} {\bibfnamefont {G.}~\bibnamefont
  {R{\'i}os}},\ }\bibfield  {title} {\bibinfo {title} {{Remarks on pole
  trajectories for resonances}},\ }\href
  {https://doi.org/10.1016/j.physletb.2014.11.011} {\bibfield  {journal}
  {\bibinfo  {journal} {Phys. Lett. B}\ }\textbf {\bibinfo {volume} {739}},\
  \bibinfo {pages} {375} (\bibinfo {year} {2014})},\ \Eprint
  {https://arxiv.org/abs/1407.7452} {arXiv:1407.7452 [hep-ph]} \BibitemShut
  {NoStop}%
\bibitem [{\citenamefont {Guo}\ and\ \citenamefont
  {Oller}(2016{\natexlab{a}})}]{Guo:2015daa}%
  \BibitemOpen
  \bibfield  {author} {\bibinfo {author} {\bibfnamefont {Z.-H.}\ \bibnamefont
  {Guo}}\ and\ \bibinfo {author} {\bibfnamefont {J.~A.}\ \bibnamefont
  {Oller}},\ }\bibfield  {title} {\bibinfo {title} {{Probabilistic
  interpretation of compositeness relation for resonances}},\ }\href
  {https://doi.org/10.1103/PhysRevD.93.096001} {\bibfield  {journal} {\bibinfo
  {journal} {Phys. Rev. D}\ }\textbf {\bibinfo {volume} {93}},\ \bibinfo
  {pages} {096001} (\bibinfo {year} {2016}{\natexlab{a}})},\ \Eprint
  {https://arxiv.org/abs/1508.06400} {arXiv:1508.06400 [hep-ph]} \BibitemShut
  {NoStop}%
\bibitem [{\citenamefont {Sekihara}\ \emph {et~al.}(2016)\citenamefont
  {Sekihara}, \citenamefont {Arai}, \citenamefont {Yamagata-Sekihara},\ and\
  \citenamefont {Yasui}}]{Sekihara:2015gvw}%
  \BibitemOpen
  \bibfield  {author} {\bibinfo {author} {\bibfnamefont {T.}~\bibnamefont
  {Sekihara}}, \bibinfo {author} {\bibfnamefont {T.}~\bibnamefont {Arai}},
  \bibinfo {author} {\bibfnamefont {J.}~\bibnamefont {Yamagata-Sekihara}},\
  and\ \bibinfo {author} {\bibfnamefont {S.}~\bibnamefont {Yasui}},\ }\bibfield
   {title} {\bibinfo {title} {{Compositeness of baryonic resonances:
  Application to the $\Delta$(1232), $N$(1535), and $N$(1650) resonances}},\
  }\href {https://doi.org/10.1103/PhysRevC.93.035204} {\bibfield  {journal}
  {\bibinfo  {journal} {Phys. Rev. C}\ }\textbf {\bibinfo {volume} {93}},\
  \bibinfo {pages} {035204} (\bibinfo {year} {2016})},\ \Eprint
  {https://arxiv.org/abs/1511.01200} {arXiv:1511.01200 [hep-ph]} \BibitemShut
  {NoStop}%
\bibitem [{\citenamefont {Kamiya}\ and\ \citenamefont
  {Hyodo}(2016)}]{Kamiya:2015aea}%
  \BibitemOpen
  \bibfield  {author} {\bibinfo {author} {\bibfnamefont {Y.}~\bibnamefont
  {Kamiya}}\ and\ \bibinfo {author} {\bibfnamefont {T.}~\bibnamefont {Hyodo}},\
  }\bibfield  {title} {\bibinfo {title} {{Structure of near-threshold
  quasibound states}},\ }\href {https://doi.org/10.1103/PhysRevC.93.035203}
  {\bibfield  {journal} {\bibinfo  {journal} {Phys. Rev. C}\ }\textbf {\bibinfo
  {volume} {93}},\ \bibinfo {pages} {035203} (\bibinfo {year} {2016})},\
  \Eprint {https://arxiv.org/abs/1509.00146} {arXiv:1509.00146 [hep-ph]}
  \BibitemShut {NoStop}%
\bibitem [{\citenamefont {Xiao}\ and\ \citenamefont
  {Zhou}(2016)}]{Xiao:2016dsx}%
  \BibitemOpen
  \bibfield  {author} {\bibinfo {author} {\bibfnamefont {Z.}~\bibnamefont
  {Xiao}}\ and\ \bibinfo {author} {\bibfnamefont {Z.-Y.}\ \bibnamefont
  {Zhou}},\ }\bibfield  {title} {\bibinfo {title} {Virtual states and
  generalized completeness relation in the {{Friedrichs Model}}},\ }\href
  {https://doi.org/10.1103/PhysRevD.94.076006} {\bibfield  {journal} {\bibinfo
  {journal} {Phys. Rev. D}\ }\textbf {\bibinfo {volume} {94}},\ \bibinfo
  {pages} {076006} (\bibinfo {year} {2016})},\ \Eprint
  {https://arxiv.org/abs/1608.00468} {arXiv:1608.00468} \BibitemShut {NoStop}%
\bibitem [{\citenamefont {Xiao}\ and\ \citenamefont
  {Zhou}(2017)}]{Xiao:2016wbs}%
  \BibitemOpen
  \bibfield  {author} {\bibinfo {author} {\bibfnamefont {Z.}~\bibnamefont
  {Xiao}}\ and\ \bibinfo {author} {\bibfnamefont {Z.-Y.}\ \bibnamefont
  {Zhou}},\ }\bibfield  {title} {\bibinfo {title} {On {{Friedrichs Model}} with
  {{Two Continuum States}}},\ }\href {https://doi.org/10.1063/1.4989832}
  {\bibfield  {journal} {\bibinfo  {journal} {J. Math. Phys.}\ }\textbf
  {\bibinfo {volume} {58}},\ \bibinfo {pages} {062110} (\bibinfo {year}
  {2017})},\ \Eprint {https://arxiv.org/abs/1608.06833} {arXiv:1608.06833}
  \BibitemShut {NoStop}%
\bibitem [{\citenamefont {Kang}\ \emph {et~al.}(2016)\citenamefont {Kang},
  \citenamefont {Guo},\ and\ \citenamefont {Oller}}]{Kang:2016ezb}%
  \BibitemOpen
  \bibfield  {author} {\bibinfo {author} {\bibfnamefont {X.-W.}\ \bibnamefont
  {Kang}}, \bibinfo {author} {\bibfnamefont {Z.-H.}\ \bibnamefont {Guo}},\ and\
  \bibinfo {author} {\bibfnamefont {J.~A.}\ \bibnamefont {Oller}},\ }\bibfield
  {title} {\bibinfo {title} {General considerations on the nature of
  ${{Z}}_b(10610)$ and ${{Z}}_b(10650)$ from their pole positions},\ }\href
  {https://doi.org/10.1103/PhysRevD.94.014012} {\bibfield  {journal} {\bibinfo
  {journal} {Phys. Rev. D}\ }\textbf {\bibinfo {volume} {94}},\ \bibinfo
  {pages} {014012} (\bibinfo {year} {2016})},\ \Eprint
  {https://arxiv.org/abs/1603.05546} {arXiv:1603.05546} \BibitemShut {NoStop}%
\bibitem [{\citenamefont {Sekihara}(2017)}]{Sekihara:2016xnq}%
  \BibitemOpen
  \bibfield  {author} {\bibinfo {author} {\bibfnamefont {T.}~\bibnamefont
  {Sekihara}},\ }\bibfield  {title} {\bibinfo {title} {{Two-body wave functions
  and compositeness from scattering amplitudes. I. General properties with
  schematic models}},\ }\href {https://doi.org/10.1103/PhysRevC.95.025206}
  {\bibfield  {journal} {\bibinfo  {journal} {Phys. Rev. C}\ }\textbf {\bibinfo
  {volume} {95}},\ \bibinfo {pages} {025206} (\bibinfo {year} {2017})},\
  \Eprint {https://arxiv.org/abs/1609.09496} {arXiv:1609.09496 [quant-ph]}
  \BibitemShut {NoStop}%
\bibitem [{\citenamefont {Kamiya}\ and\ \citenamefont
  {Hyodo}(2017{\natexlab{a}})}]{Kamiya:2016oao}%
  \BibitemOpen
  \bibfield  {author} {\bibinfo {author} {\bibfnamefont {Y.}~\bibnamefont
  {Kamiya}}\ and\ \bibinfo {author} {\bibfnamefont {T.}~\bibnamefont {Hyodo}},\
  }\bibfield  {title} {\bibinfo {title} {{Generalized weak-binding relations of
  compositeness in effective field theory}},\ }\href
  {https://doi.org/10.1093/ptep/ptw188} {\bibfield  {journal} {\bibinfo
  {journal} {PTEP}\ }\textbf {\bibinfo {volume} {2017}},\ \bibinfo {pages}
  {023D02} (\bibinfo {year} {2017}{\natexlab{a}})},\ \Eprint
  {https://arxiv.org/abs/1607.01899} {arXiv:1607.01899 [hep-ph]} \BibitemShut
  {NoStop}%
\bibitem [{\citenamefont {Guo}\ and\ \citenamefont
  {Oller}(2016{\natexlab{b}})}]{Guo:2016wpy}%
  \BibitemOpen
  \bibfield  {author} {\bibinfo {author} {\bibfnamefont {Z.-H.}\ \bibnamefont
  {Guo}}\ and\ \bibinfo {author} {\bibfnamefont {J.~A.}\ \bibnamefont
  {Oller}},\ }\bibfield  {title} {\bibinfo {title} {Resonance on top of
  thresholds: The ${\Lambda}_c(2595)^+$ as an extremely fine-tuned state},\
  }\href {https://doi.org/10.1103/PhysRevD.93.054014} {\bibfield  {journal}
  {\bibinfo  {journal} {Phys. Rev. D}\ }\textbf {\bibinfo {volume} {93}},\
  \bibinfo {pages} {054014} (\bibinfo {year} {2016}{\natexlab{b}})},\ \Eprint
  {https://arxiv.org/abs/1601.00862} {arXiv:1601.00862} \BibitemShut {NoStop}%
\bibitem [{\citenamefont {Oller}(2018)}]{Oller:2017alp}%
  \BibitemOpen
  \bibfield  {author} {\bibinfo {author} {\bibfnamefont {J.~A.}\ \bibnamefont
  {Oller}},\ }\bibfield  {title} {\bibinfo {title} {{New results from a number
  operator interpretation of the compositeness of bound and resonant states}},\
  }\href {https://doi.org/10.1016/j.aop.2018.07.023} {\bibfield  {journal}
  {\bibinfo  {journal} {Annals Phys.}\ }\textbf {\bibinfo {volume} {396}},\
  \bibinfo {pages} {429} (\bibinfo {year} {2018})},\ \Eprint
  {https://arxiv.org/abs/1710.00991} {arXiv:1710.00991 [hep-ph]} \BibitemShut
  {NoStop}%
\bibitem [{\citenamefont {Kamiya}\ and\ \citenamefont
  {Hyodo}(2017{\natexlab{b}})}]{Kamiya:2017hni}%
  \BibitemOpen
  \bibfield  {author} {\bibinfo {author} {\bibfnamefont {Y.}~\bibnamefont
  {Kamiya}}\ and\ \bibinfo {author} {\bibfnamefont {T.}~\bibnamefont {Hyodo}},\
  }\bibfield  {title} {\bibinfo {title} {{Compositeness of Quasibound States
  from Effective Field Theory}},\ }\href {https://doi.org/10.22323/1.281.0270}
  {\bibfield  {journal} {\bibinfo  {journal} {PoS}\ }\textbf {\bibinfo {volume}
  {INPC2016}},\ \bibinfo {pages} {270} (\bibinfo {year}
  {2017}{\natexlab{b}})},\ \Eprint {https://arxiv.org/abs/1701.08941}
  {arXiv:1701.08941 [hep-ph]} \BibitemShut {NoStop}%
\bibitem [{\citenamefont {Bruns}(2019)}]{Bruns:2019xgo}%
  \BibitemOpen
  \bibfield  {author} {\bibinfo {author} {\bibfnamefont {P.~C.}\ \bibnamefont
  {Bruns}},\ }\bibfield  {title} {\bibinfo {title} {Spatial interpretation of
  ``compositeness" for finite-range potentials},\ }\href@noop {} {\bibfield
  {journal} {\bibinfo  {journal} {arXiv:1905.09196 [hep-ph]}\ } (\bibinfo
  {year} {2019})},\ \Eprint {https://arxiv.org/abs/1905.09196}
  {arXiv:1905.09196 [hep-ph]} \BibitemShut {NoStop}%
\bibitem [{\citenamefont {Matuschek}\ \emph {et~al.}(2021)\citenamefont
  {Matuschek}, \citenamefont {Baru}, \citenamefont {Guo},\ and\ \citenamefont
  {Hanhart}}]{Matuschek:2020gqe}%
  \BibitemOpen
  \bibfield  {author} {\bibinfo {author} {\bibfnamefont {I.}~\bibnamefont
  {Matuschek}}, \bibinfo {author} {\bibfnamefont {V.}~\bibnamefont {Baru}},
  \bibinfo {author} {\bibfnamefont {F.-K.}\ \bibnamefont {Guo}},\ and\ \bibinfo
  {author} {\bibfnamefont {C.}~\bibnamefont {Hanhart}},\ }\bibfield  {title}
  {\bibinfo {title} {{On the nature of near-threshold bound and virtual
  states}},\ }\href {https://doi.org/10.1140/epja/s10050-021-00413-y}
  {\bibfield  {journal} {\bibinfo  {journal} {Eur. Phys. J. A}\ }\textbf
  {\bibinfo {volume} {57}},\ \bibinfo {pages} {101} (\bibinfo {year} {2021})},\
  \Eprint {https://arxiv.org/abs/2007.05329} {arXiv:2007.05329 [hep-ph]}
  \BibitemShut {NoStop}%
\bibitem [{\citenamefont {Guo}\ \emph {et~al.}(2018)\citenamefont {Guo},
  \citenamefont {Hanhart}, \citenamefont {Mei\ss{}ner}, \citenamefont {Wang},
  \citenamefont {Zhao},\ and\ \citenamefont {Zou}}]{Guo:2017jvc}%
  \BibitemOpen
  \bibfield  {author} {\bibinfo {author} {\bibfnamefont {F.-K.}\ \bibnamefont
  {Guo}}, \bibinfo {author} {\bibfnamefont {C.}~\bibnamefont {Hanhart}},
  \bibinfo {author} {\bibfnamefont {U.-G.}\ \bibnamefont {Mei\ss{}ner}},
  \bibinfo {author} {\bibfnamefont {Q.}~\bibnamefont {Wang}}, \bibinfo {author}
  {\bibfnamefont {Q.}~\bibnamefont {Zhao}},\ and\ \bibinfo {author}
  {\bibfnamefont {B.-S.}\ \bibnamefont {Zou}},\ }\bibfield  {title} {\bibinfo
  {title} {{Hadronic molecules}},\ }\href
  {https://doi.org/10.1103/RevModPhys.90.015004} {\bibfield  {journal}
  {\bibinfo  {journal} {Rev. Mod. Phys.}\ }\textbf {\bibinfo {volume} {90}},\
  \bibinfo {pages} {015004} (\bibinfo {year} {2018})},\ \Eprint
  {https://arxiv.org/abs/1705.00141} {arXiv:1705.00141 [hep-ph]} \BibitemShut
  {NoStop}%
\bibitem [{\citenamefont {Zhaba}(2017)}]{Zhaba:2017syr}%
  \BibitemOpen
  \bibfield  {author} {\bibinfo {author} {\bibfnamefont {V.~I.}\ \bibnamefont
  {Zhaba}},\ }\bibfield  {title} {\bibinfo {title} {{Deuteron: properties and
  analytical forms of wave function in coordinate space}},\ }\href@noop {}
  {\bibfield  {journal} {\bibinfo  {journal} {arXiv:1706.08306 [nucl-th]}\ }
  (\bibinfo {year} {2017})},\ \Eprint {https://arxiv.org/abs/1706.08306}
  {arXiv:1706.08306 [nucl-th]} \BibitemShut {NoStop}%
\bibitem [{\citenamefont {Chadan}\ \emph {et~al.}(1989)\citenamefont {Chadan},
  \citenamefont {Sabatier},\ and\ \citenamefont {Newton}}]{Chadan:1989Inverse}%
  \BibitemOpen
  \bibfield  {author} {\bibinfo {author} {\bibfnamefont {K.}~\bibnamefont
  {Chadan}}, \bibinfo {author} {\bibfnamefont {P.~C.}\ \bibnamefont
  {Sabatier}},\ and\ \bibinfo {author} {\bibfnamefont {R.~G.}\ \bibnamefont
  {Newton}},\ }\href {https://doi.org/10.1007/978-3-642-83317-5} {\emph
  {\bibinfo {title} {Inverse {{Problems}} in {{Quantum Scattering Theory}}}}}\
  (\bibinfo  {publisher} {{Springer Berlin Heidelberg}},\ \bibinfo {address}
  {{Berlin, Heidelberg}},\ \bibinfo {year} {1989})\BibitemShut {NoStop}%
\bibitem [{\citenamefont {Castillejo}\ \emph {et~al.}(1956)\citenamefont
  {Castillejo}, \citenamefont {Dalitz},\ and\ \citenamefont
  {Dyson}}]{Castillejo:1955ed}%
  \BibitemOpen
  \bibfield  {author} {\bibinfo {author} {\bibfnamefont {L.}~\bibnamefont
  {Castillejo}}, \bibinfo {author} {\bibfnamefont {R.~H.}\ \bibnamefont
  {Dalitz}},\ and\ \bibinfo {author} {\bibfnamefont {F.~J.}\ \bibnamefont
  {Dyson}},\ }\bibfield  {title} {\bibinfo {title} {{Low's scattering equation
  for the charged and neutral scalar theories}},\ }\href
  {https://doi.org/10.1103/PhysRev.101.453} {\bibfield  {journal} {\bibinfo
  {journal} {Phys. Rev.}\ }\textbf {\bibinfo {volume} {101}},\ \bibinfo {pages}
  {453} (\bibinfo {year} {1956})}\BibitemShut {NoStop}%
\bibitem [{\citenamefont {Kang}\ and\ \citenamefont
  {Oller}(2017)}]{Kang:2016jxw}%
  \BibitemOpen
  \bibfield  {author} {\bibinfo {author} {\bibfnamefont {X.-W.}\ \bibnamefont
  {Kang}}\ and\ \bibinfo {author} {\bibfnamefont {J.~A.}\ \bibnamefont
  {Oller}},\ }\bibfield  {title} {\bibinfo {title} {{Different pole structures
  in line shapes of the $X(3872)$}},\ }\href
  {https://doi.org/10.1140/epjc/s10052-017-4961-z} {\bibfield  {journal}
  {\bibinfo  {journal} {Eur. Phys. J. C}\ }\textbf {\bibinfo {volume} {77}},\
  \bibinfo {pages} {399} (\bibinfo {year} {2017})},\ \Eprint
  {https://arxiv.org/abs/1612.08420} {arXiv:1612.08420 [hep-ph]} \BibitemShut
  {NoStop}%
\bibitem [{\citenamefont {Faessler}\ \emph {et~al.}(2007)\citenamefont
  {Faessler}, \citenamefont {Gutsche}, \citenamefont {Lyubovitskij},\ and\
  \citenamefont {Ma}}]{Faessler:2007gv}%
  \BibitemOpen
  \bibfield  {author} {\bibinfo {author} {\bibfnamefont {A.}~\bibnamefont
  {Faessler}}, \bibinfo {author} {\bibfnamefont {T.}~\bibnamefont {Gutsche}},
  \bibinfo {author} {\bibfnamefont {V.~E.}\ \bibnamefont {Lyubovitskij}},\ and\
  \bibinfo {author} {\bibfnamefont {Y.-L.}\ \bibnamefont {Ma}},\ }\bibfield
  {title} {\bibinfo {title} {{Strong and radiative decays of the
  $D_{s0}^*(2317)$ meson in the $DK$-molecule picture}},\ }\href
  {https://doi.org/10.1103/PhysRevD.76.014005} {\bibfield  {journal} {\bibinfo
  {journal} {Phys. Rev. D}\ }\textbf {\bibinfo {volume} {76}},\ \bibinfo
  {pages} {014005} (\bibinfo {year} {2007})},\ \Eprint
  {https://arxiv.org/abs/0705.0254} {arXiv:0705.0254 [hep-ph]} \BibitemShut
  {NoStop}%
\bibitem [{\citenamefont {Bali}\ \emph {et~al.}(2017)\citenamefont {Bali},
  \citenamefont {Collins}, \citenamefont {Cox},\ and\ \citenamefont
  {Sch\"afer}}]{Bali:2017pdv}%
  \BibitemOpen
  \bibfield  {author} {\bibinfo {author} {\bibfnamefont {G.~S.}\ \bibnamefont
  {Bali}}, \bibinfo {author} {\bibfnamefont {S.}~\bibnamefont {Collins}},
  \bibinfo {author} {\bibfnamefont {A.}~\bibnamefont {Cox}},\ and\ \bibinfo
  {author} {\bibfnamefont {A.}~\bibnamefont {Sch\"afer}},\ }\bibfield  {title}
  {\bibinfo {title} {{Masses and decay constants of the $D_{s0}^*(2317)$ and
  $D_{s1}(2460)$ from $N_f=2$ lattice QCD close to the physical point}},\
  }\href {https://doi.org/10.1103/PhysRevD.96.074501} {\bibfield  {journal}
  {\bibinfo  {journal} {Phys. Rev. D}\ }\textbf {\bibinfo {volume} {96}},\
  \bibinfo {pages} {074501} (\bibinfo {year} {2017})},\ \Eprint
  {https://arxiv.org/abs/1706.01247} {arXiv:1706.01247 [hep-lat]} \BibitemShut
  {NoStop}%
\bibitem [{\citenamefont {Cheung}\ \emph {et~al.}(2021)\citenamefont {Cheung},
  \citenamefont {Thomas}, \citenamefont {Wilson}, \citenamefont {Moir},
  \citenamefont {Peardon},\ and\ \citenamefont {Ryan}}]{Cheung:2020mql}%
  \BibitemOpen
  \bibfield  {author} {\bibinfo {author} {\bibfnamefont {G.~K.~C.}\
  \bibnamefont {Cheung}}, \bibinfo {author} {\bibfnamefont {C.~E.}\
  \bibnamefont {Thomas}}, \bibinfo {author} {\bibfnamefont {D.~J.}\
  \bibnamefont {Wilson}}, \bibinfo {author} {\bibfnamefont {G.}~\bibnamefont
  {Moir}}, \bibinfo {author} {\bibfnamefont {M.}~\bibnamefont {Peardon}},\ and\
  \bibinfo {author} {\bibfnamefont {S.~M.}\ \bibnamefont {Ryan}} (\bibinfo
  {collaboration} {Hadron Spectrum}),\ }\bibfield  {title} {\bibinfo {title}
  {{$DK$ $I = 0$, $ D\overline{K} $ $I = 0, 1$ scattering and the $
  {D}_{s0}^{\ast } $(2317) from lattice QCD}},\ }\href
  {https://doi.org/10.1007/JHEP02(2021)100} {\bibfield  {journal} {\bibinfo
  {journal} {JHEP}\ }\textbf {\bibinfo {volume} {02}},\ \bibinfo {pages}
  {100}},\ \Eprint {https://arxiv.org/abs/2008.06432} {arXiv:2008.06432
  [hep-lat]} \BibitemShut {NoStop}%
\bibitem [{\citenamefont {Bohm}(2001)}]{Bohm:book}%
  \BibitemOpen
  \bibfield  {author} {\bibinfo {author} {\bibfnamefont {A.}~\bibnamefont
  {Bohm}},\ }\href@noop {} {\emph {\bibinfo {title} {Quantum mechanics:
  foundations and applications, 3rd edn.}}}\ (\bibinfo  {publisher} {Springer,
  Berlin},\ \bibinfo {year} {2001})\BibitemShut {NoStop}%
\bibitem [{\citenamefont {Hanhart}\ \emph {et~al.}(2007)\citenamefont
  {Hanhart}, \citenamefont {Kalashnikova}, \citenamefont {Kudryavtsev},\ and\
  \citenamefont {Nefediev}}]{Hanhart:2007wa}%
  \BibitemOpen
  \bibfield  {author} {\bibinfo {author} {\bibfnamefont {C.}~\bibnamefont
  {Hanhart}}, \bibinfo {author} {\bibfnamefont {Y.~S.}\ \bibnamefont
  {Kalashnikova}}, \bibinfo {author} {\bibfnamefont {A.~E.}\ \bibnamefont
  {Kudryavtsev}},\ and\ \bibinfo {author} {\bibfnamefont {A.~V.}\ \bibnamefont
  {Nefediev}},\ }\bibfield  {title} {\bibinfo {title} {{Two-photon decays of
  hadronic molecules}},\ }\href {https://doi.org/10.1103/PhysRevD.75.074015}
  {\bibfield  {journal} {\bibinfo  {journal} {Phys. Rev. D}\ }\textbf {\bibinfo
  {volume} {75}},\ \bibinfo {pages} {074015} (\bibinfo {year} {2007})},\
  \Eprint {https://arxiv.org/abs/hep-ph/0701214} {arXiv:hep-ph/0701214}
  \BibitemShut {NoStop}%
\bibitem [{\citenamefont {Stoks}\ \emph {et~al.}(1994)\citenamefont {Stoks},
  \citenamefont {Klomp}, \citenamefont {Terheggen},\ and\ \citenamefont
  {de~Swart}}]{Stoks:1994wp}%
  \BibitemOpen
  \bibfield  {author} {\bibinfo {author} {\bibfnamefont {V.~G.~J.}\
  \bibnamefont {Stoks}}, \bibinfo {author} {\bibfnamefont {R.~A.~M.}\
  \bibnamefont {Klomp}}, \bibinfo {author} {\bibfnamefont {C.~P.~F.}\
  \bibnamefont {Terheggen}},\ and\ \bibinfo {author} {\bibfnamefont {J.~J.}\
  \bibnamefont {de~Swart}},\ }\bibfield  {title} {\bibinfo {title}
  {{Construction of high quality N N potential models}},\ }\href
  {https://doi.org/10.1103/PhysRevC.49.2950} {\bibfield  {journal} {\bibinfo
  {journal} {Phys. Rev. C}\ }\textbf {\bibinfo {volume} {49}},\ \bibinfo
  {pages} {2950} (\bibinfo {year} {1994})},\ \Eprint
  {https://arxiv.org/abs/nucl-th/9406039} {arXiv:nucl-th/9406039} \BibitemShut
  {NoStop}%
\bibitem [{nno()}]{nnonline}%
  \BibitemOpen
  \href@noop {} {}\bibinfo {howpublished}
  {\url{http://nn-online.org}}\BibitemShut {NoStop}%
\bibitem [{\citenamefont {Wiringa}\ \emph {et~al.}(1995)\citenamefont
  {Wiringa}, \citenamefont {Stoks},\ and\ \citenamefont
  {Schiavilla}}]{Wiringa:1994wb}%
  \BibitemOpen
  \bibfield  {author} {\bibinfo {author} {\bibfnamefont {R.~B.}\ \bibnamefont
  {Wiringa}}, \bibinfo {author} {\bibfnamefont {V.~G.~J.}\ \bibnamefont
  {Stoks}},\ and\ \bibinfo {author} {\bibfnamefont {R.}~\bibnamefont
  {Schiavilla}},\ }\bibfield  {title} {\bibinfo {title} {{An accurate
  nucleon-nucleon potential with charge independence breaking}},\ }\href
  {https://doi.org/10.1103/PhysRevC.51.38} {\bibfield  {journal} {\bibinfo
  {journal} {Phys. Rev. C}\ }\textbf {\bibinfo {volume} {51}},\ \bibinfo
  {pages} {38} (\bibinfo {year} {1995})},\ \Eprint
  {https://arxiv.org/abs/nucl-th/9408016} {arXiv:nucl-th/9408016} \BibitemShut
  {NoStop}%
\bibitem [{\citenamefont {Epelbaum}\ \emph {et~al.}(2015)\citenamefont
  {Epelbaum}, \citenamefont {Krebs},\ and\ \citenamefont
  {Mei\ss{}ner}}]{Epelbaum:2014efa}%
  \BibitemOpen
  \bibfield  {author} {\bibinfo {author} {\bibfnamefont {E.}~\bibnamefont
  {Epelbaum}}, \bibinfo {author} {\bibfnamefont {H.}~\bibnamefont {Krebs}},\
  and\ \bibinfo {author} {\bibfnamefont {U.-G.}\ \bibnamefont {Mei\ss{}ner}},\
  }\bibfield  {title} {\bibinfo {title} {{Improved chiral nucleon-nucleon
  potential up to next-to-next-to-next-to-leading order}},\ }\href
  {https://doi.org/10.1140/epja/i2015-15053-8} {\bibfield  {journal} {\bibinfo
  {journal} {Eur. Phys. J. A}\ }\textbf {\bibinfo {volume} {51}},\ \bibinfo
  {pages} {53} (\bibinfo {year} {2015})},\ \Eprint
  {https://arxiv.org/abs/1412.0142} {arXiv:1412.0142 [nucl-th]} \BibitemShut
  {NoStop}%
\bibitem [{\citenamefont {Entem}\ and\ \citenamefont
  {Machleidt}(2003)}]{Entem:2003ft}%
  \BibitemOpen
  \bibfield  {author} {\bibinfo {author} {\bibfnamefont {D.~R.}\ \bibnamefont
  {Entem}}\ and\ \bibinfo {author} {\bibfnamefont {R.}~\bibnamefont
  {Machleidt}},\ }\bibfield  {title} {\bibinfo {title} {{Accurate charge
  dependent nucleon nucleon potential at fourth order of chiral perturbation
  theory}},\ }\href {https://doi.org/10.1103/PhysRevC.68.041001} {\bibfield
  {journal} {\bibinfo  {journal} {Phys. Rev. C}\ }\textbf {\bibinfo {volume}
  {68}},\ \bibinfo {pages} {041001} (\bibinfo {year} {2003})},\ \Eprint
  {https://arxiv.org/abs/nucl-th/0304018} {arXiv:nucl-th/0304018} \BibitemShut
  {NoStop}%
\bibitem [{\citenamefont {Epelbaum}\ \emph {et~al.}(2005)\citenamefont
  {Epelbaum}, \citenamefont {Glockle},\ and\ \citenamefont
  {Mei{\ss}ner}}]{Epelbaum:2004fk}%
  \BibitemOpen
  \bibfield  {author} {\bibinfo {author} {\bibfnamefont {E.}~\bibnamefont
  {Epelbaum}}, \bibinfo {author} {\bibfnamefont {W.}~\bibnamefont {Glockle}},\
  and\ \bibinfo {author} {\bibfnamefont {U.-G.}\ \bibnamefont {Mei{\ss}ner}},\
  }\bibfield  {title} {\bibinfo {title} {{The two-nucleon system at
  next-to-next-to-next-to-leading order}},\ }\href
  {https://doi.org/10.1016/j.nuclphysa.2004.09.107} {\bibfield  {journal}
  {\bibinfo  {journal} {Nucl. Phys. A}\ }\textbf {\bibinfo {volume} {747}},\
  \bibinfo {pages} {362} (\bibinfo {year} {2005})},\ \Eprint
  {https://arxiv.org/abs/nucl-th/0405048} {arXiv:nucl-th/0405048} \BibitemShut
  {NoStop}%
\bibitem [{\citenamefont {Machleidt}(2001)}]{Machleidt:2000ge}%
  \BibitemOpen
  \bibfield  {author} {\bibinfo {author} {\bibfnamefont {R.}~\bibnamefont
  {Machleidt}},\ }\bibfield  {title} {\bibinfo {title} {{The high precision,
  charge dependent Bonn nucleon-nucleon potential (CD-Bonn)}},\ }\href
  {https://doi.org/10.1103/PhysRevC.63.024001} {\bibfield  {journal} {\bibinfo
  {journal} {Phys. Rev. C}\ }\textbf {\bibinfo {volume} {63}},\ \bibinfo
  {pages} {024001} (\bibinfo {year} {2001})},\ \Eprint
  {https://arxiv.org/abs/nucl-th/0006014} {arXiv:nucl-th/0006014} \BibitemShut
  {NoStop}%
\bibitem [{\citenamefont {Nogga}\ and\ \citenamefont
  {Hanhart}(2006)}]{Nogga:2005fv}%
  \BibitemOpen
  \bibfield  {author} {\bibinfo {author} {\bibfnamefont {A.}~\bibnamefont
  {Nogga}}\ and\ \bibinfo {author} {\bibfnamefont {C.}~\bibnamefont
  {Hanhart}},\ }\bibfield  {title} {\bibinfo {title} {Can one extract the
  $\pi$-neutron scattering length from $\pi$-deuteron scattering?},\ }\href
  {https://doi.org/10.1016/j.physletb.2005.12.071} {\bibfield  {journal}
  {\bibinfo  {journal} {Physics Letters B}\ }\textbf {\bibinfo {volume}
  {634}},\ \bibinfo {pages} {210} (\bibinfo {year} {2006})},\ \Eprint
  {https://arxiv.org/abs/nucl-th/0511011} {arXiv:nucl-th/0511011} \BibitemShut
  {NoStop}%
\bibitem [{\citenamefont {Babenko}\ and\ \citenamefont
  {Petrov}(2005)}]{Babenko:2005qp}%
  \BibitemOpen
  \bibfield  {author} {\bibinfo {author} {\bibfnamefont {V.~A.}\ \bibnamefont
  {Babenko}}\ and\ \bibinfo {author} {\bibfnamefont {N.~M.}\ \bibnamefont
  {Petrov}},\ }\bibfield  {title} {\bibinfo {title} {{Description of the
  two-nucleon system on the basis of the Bargmann representation of the $S$
  matrix}},\ }\href {https://doi.org/10.1134/1.1866377} {\bibfield  {journal}
  {\bibinfo  {journal} {Phys. Atom. Nucl.}\ }\textbf {\bibinfo {volume} {68}},\
  \bibinfo {pages} {219} (\bibinfo {year} {2005})},\ \Eprint
  {https://arxiv.org/abs/nucl-th/0502041} {arXiv:nucl-th/0502041} \BibitemShut
  {NoStop}%
\end{thebibliography}%

\onecolumngrid
\newpage
\section*{Supplemental Material}
\setcounter{equation}{0}

\subsection{Solution of the Low equation}
The Low equation with $V_{p,k}$ neglected reads
\begin{align}
    T_{p,k} = \frac{g(p)\, g^*(k)}{h_k-E_B } + \int_0^\infty \frac{q^2dq}{(2\pi)^3} \frac{T_{p,q}T_{k,q}^*}{h_k+i\varepsilon-h_q} \,.
    \label{eq:appT}
\end{align}
As the only $T$-independent term of this equation is of a separable form, it is reasonable to consider also a separable ansatz:
\begin{align}
    T_{p,k} = t_k\, g(p)\, g^*(k) \,.
\end{align}
Then the Low equation becomes
\begin{align}
    t_k = \frac{1}{h_k - E_B} + \int_0^\infty \frac{q^2dq}{(2\pi)^3} \frac{|t_q|^2|g(q)|^2}{h_k+i\varepsilon-h_q} \,.
\end{align}
We also define the analytic continuation of $t_k$ as
\begin{align}
    \tau(W) :=  \frac{1}{W - E_B} + \int_0^\infty \frac{q^2dq}{(2\pi)^3} \frac{|t_q|^2|g(q)|^2}{W-h_q} \,.
\end{align}

Now, it is better to work with $\tau^{-1}(W)$ because we have
\begin{align}
    \mathrm{Im}\, \tau^{-1}(h_p+i\varepsilon) 
    = \frac{\pi\, p\,\mu}{(2\pi)^3} |g(p)|^2 \theta(h_p)\,,
\end{align}
as a consequence of unitarity, where we have used $\tau(h_p+i\varepsilon)=t_p$.
$\tau^{-1}(W)$ is a real analytic function with possible singularities residing only on the real axis.
These singularities correspond to zeros of $\tau(W)$,
which are known as the Castillejo-Dalitz-Dyson (CDD) zeros~\cite{Castillejo:1955ed}.
As Weinberg did, we will look for a solution without such zeros (for a discussion of the impact of CDD zeros on the compositeness, see Refs.~\cite{Baru:2010ww,Hanhart:2011jz,Kang:2016jxw}).
Then a twice-subtracted dispersive relation gives
\begin{align}
    \tau^{-1}(W) 
    =\, (W-E_B) 
    + (W-E_B)^2 \int_0^\infty \frac{q^2dq}{(2\pi)^3} \frac{|g(q)|^2}{(h_q-E_B)^2(h_q-W)} \,,
\end{align}
where we have used
\begin{align}
    \tau^{-1}(E_B) = 0 \,,\qquad \tau^{-1'}(E_B) = 1 \,.
\end{align}
Finally, one gets
\begin{align}
    \tau(W) = \frac{1}{1-F(W)}\,\frac{1}{W-E_B}, 
\end{align}
with
\begin{align}
    F(W) &:=  (W-E_B) \int_0^\infty \frac{q^2dq}{(2\pi)^3} \frac{|g(q)|^2}{(h_q-E_B)^2(W-h_q)} \,.
\end{align}

One can further work out a dispersive representation for the function $F(W)$. For that, it is convenient to define
\begin{align}
    F_1(W):= \frac{\ln \left[1-F(W)\right]}{W-E_B}\,.
\end{align}
When $E>0$, one has
\begin{align}
    \mathrm{Im}\,F_1(E+i\varepsilon)  = - \frac{\delta_B(E)}{E-E_B} \,
\end{align}
with $\delta_B$ the phase of the on-shell $T$-matrix $T_{k,k}$ as given by the solution of Eq.~\eqref{eq:appT}; see Eq.~\eqref{eq:argdelta} in the main text.
With the convention $\delta_B(0)=0$ we have taken, one can show that $F_1(E)$ is real for $E\leq0$ by noticing that $F(E)$ is monotonically increasing in the same region and $F(0)\leq0$.
So it is a real analytic function, allowing for a standard dispersion relation which gives
\begin{align}
    F_1(W) = -\frac{1}{\pi}\int_0^\infty dE \frac{\delta_B(E)}{(E-W)(E-E_B)} \,.
\end{align}
Then one gets
\begin{align}
    F(W) = 1 - \exp\left(\frac{W-E_B}{\pi}\int_0^\infty dE \frac{-\delta_B(E)}{(E-W)(E-E_B)}\right) \,.
\end{align}

\end{document}